\documentclass[a4paper,fleqn,usenatbib,useAMS]{mnras}
\usepackage[T1]{fontenc}
\usepackage{ae,aecompl}
\usepackage{graphicx}
\usepackage{amsmath}
\usepackage{amssymb}
\usepackage{lineno}

\newcommand\suzaku{{\it Suzaku}}
\newcommand\xmm{{\it XMM-Newton}}
\newcommand\rosat{{\it ROSAT}}
\newcommand\astr{{\it ASTROSAT}}

\newcommand\s{{\rm~s}}
\newcommand\ks{{\rm~ks}}

\newcommand\kpc{{\rm~kpc}}

\newcommand\hz{{\rm~Hz}}
\newcommand\ghz{{\rm~GHz}}
\newcommand\keV{{\rm~keV}}
\newcommand\ev{{\rm~eV}}

\title[Disc-Corona and Jet emission from RXJ1633.3+4719]{Accretion disc-corona and jet emission from the radio-loud Narrow Line Seyfert 1 galaxy RXJ1633.3+4719}
\author[L. Mallick et al.]{Labani Mallick$^{1,2}$\thanks{E-mail: labani@iucaa.in (LM), lmphy03@gmail.com}, G. C. Dewangan$^{1}$, P. Gandhi$^{2}$, R. Misra$^{1}$ and A. K. Kembhavi $^{1}$ \\
$^1$ Inter-University Centre for Astronomy and Astrophysics, Post Bag 4, Ganeshkhind, Pune 411007, India\\
$^2$ School of Physics \& Astronomy, University of Southampton, Highfield, Southampton SO17 1BJ, UK}

\begin{document}


\pagerange{\pageref{firstpage}--\pageref{lastpage}} 

\maketitle

\label{firstpage}

\begin{abstract}
We perform X-ray/UV spectral and X-ray variability studies of the radio-loud Narrow Line Seyfert 1 (NLS1) galaxy RXJ1633.3+4719 using \xmm{} and \suzaku{} observations from 2011 and 2012. The 0.3$-$10\keV{} spectra consist of an ultra-soft component described by an accretion disc blackbody ($kT_{\rm in} = 39.6^{+11.2}_{-5.5}$\ev{}) and a power-law due to the thermal Comptonization ($\Gamma=1.96^{+0.24}_{-0.31}$) of the disc emission. The disc temperature inferred from the soft excess is at least a factor of two lower than that found for the canonical soft excess emission from radio-quiet NLS1s. The UV spectrum is described by a power-law with photon index $3.05^{+0.56}_{-0.33}$. The observed UV emission is too strong to arise from the accretion disc or the host galaxy, but can be attributed to a jet. The X-ray emission from RXJ1633.3+4719 is variable with fractional variability amplitude $F_{\rm var}$=13.5$\pm1.0\%$. In contrast to radio-quiet AGN, X-ray emission from the source becomes harder with increasing flux. The fractional RMS variability increases with energy and the RMS spectrum is well described by a constant disc component and a variable power-law continuum with the normalization and photon index being anti-correlated. Such spectral variability cannot be caused by variations in the absorption and must be intrinsic to the hot corona. Our finding of possible evidence for emission from the inner accretion disc, jet and hot corona from RXJ1633.3+4719 in the optical to X-ray bands makes this object an ideal target to probe the disc-jet connection in AGN.  
\end{abstract}


\section{Introduction}
Active Galactic Nuclei (AGN) are the most luminous steady objects in the universe with power output ranging from $10^{41}-10^{47}$erg~s$^{-1}$. The spectra of AGN generally consist of the following components: a big blue bump, a primary X-ray continuum with a high energy cut-off, a soft X-ray excess, blurred reflection and absorption (neutral or ionized). The big blue bump is thought to arise from the accretion disc. The thermal emission from the standard accretion disc is a combination of blackbody radiation with different temperatures that depend upon the size of the emitting region $T(r)\propto r^{-3/4}$, mass accretion rate $T(r)\propto\dot{m}^{1/4}$ and the central supermassive black hole (SMBH) mass $T(r)\propto M^{-1/4}$. The peak emission from the standard accretion disc in nearby AGN falls in the far or extreme UV band \citep{sh73} and is generally not observable due to the photoelectric absorption in our Galaxy as well as in the host galaxy. This has caused a major obstacle in our understanding of the nature of the accretion disc in AGN. It is not clearly known if AGN host the standard geometrically thin and optically thick disc as modelled by \citet{sh73}. The primary X-ray emission is believed to be produced by inverse Compton scattering of the optical/UV disc photons (see e.g. \citealt{su80}). It has variously been proposed that the scattering region could be a hot corona above the cold disc \citep{ha91,ha93,po96}, an advection dominated accretion flow (ADAF) \citep{na94, es97} or a relativistic jet or base of the jet \citep{fe99,fe04,ma05}. The geometry of the corona could be of the form of a `slab' or `sandwich' \citep{po97b,hu95}, `sphere+disc geometries' \citep{do97b, po97a, gi97, ha91, ha93, ha97} or it could be a `patchy corona' \citep{st95}. A large fraction of AGN show an excess emission over the power-law continuum below $\sim2$\keV{}. The origin of this soft excess ($\sim0.1-2$\keV{}) is not clear. It may arise due to the thermal Comptonization of the disc photons in a low temperature optically thick medium \citep{ma98,de07,done12} or blurred reflection from the ionized accretion disc \citep{fa02, cr06, ga14}. The broad iron K$_\alpha$ line ($\sim4-7$\keV{}) and a Compton reflection hump ($\sim20-40$\keV{}) are the main ingredients of the blurred reflection component. The nuclear emission is generally affected by neutral or ionized absorbers that may cover the nuclear source fully or partially and may be outflowing, such as the winds from the accretion disc or `pc-scale torus' \citep{an93}. 

Narrow-Line Seyfert 1 (NLS1s) galaxies are a subclass of Seyfert 1 galaxies with relatively narrow Balmer lines FWHM(H$_\beta$) $<2000$\rm~km~s$^{-1}$ \citep{os85,go89}, strong permitted optical/UV Fe II emission lines \citep{bo92,gr99,ve01}, weaker [OIII] emission $\frac{[OIII]\lambda5007}{H_{\beta}}\leq3$ \citep{os85,go89}, steep soft X-ray spectra \citep{wa96,bo96,gr98}, rapid X-ray variability \citep{le99,ko00}, and smaller black hole masses \citep{zh06,ry07} which partially accounts for their high Eddington ratios $R_{\rm Edd}\equiv\frac{L_{\rm bol}}{L_{\rm Edd}}\approx1$ \citep{bo02,co04}. By virtue of their smaller black hole masses and high accretion rates, NLS1s are useful laboratories of an earlier stage of black hole mass build-up in AGN, also corroborated by the higher incidence of `pseudo-bulges' in their host galaxies \citep{or11}. 

In the spectral energy distributions (SED) of AGN ranging from radio to gamma rays, the disc emission usually peaks in the optical/UV regime. Radio-loud AGN generally exhibit collimated jets, sometimes dominating the emission from the disc/corona. The fundamental plane of black hole activity suggests that there exists a correlation between jet power, disc power and black hole mass \citep{ma03}. Therefore physical processes governing the disc-accretion and the jet launching mechanism are universal across mass scales. Although there is a phenomenological understanding of the disc-jet paradigm in black-hole X-ray binaries, the disc-jet connection is not well established for AGN. Despite many efforts to understand radiative processes in large scale radio jets (\citealt{mo13, do13, pa15, sc15}), the role of the innermost disc/corona region for the ejection of jets is not clearly understood in AGN. The aim of this work is to disentangle the disc/corona and jet emission and characterise the accretion disc emission by modelling the multiband (optical/UV and X-ray) data. The radio-loud Narrow Line Seyfert 1 (RLNLS1) galaxies are similar to the radio-quiet NLS1s except for their radio-loudness and are useful laboratories to study the formation of jets and their connection with the disc/corona. 

RXJ1633.3+4719 is a flat spectrum radio-loud AGN at redshift $z=0.116$ with radio-loudness parameter $R_{5}\equiv f_{\nu}$(5\ghz{})$/f_{\nu}$($4400\textrm{\AA}$)$> 100$ and is optically classified as a NLS1 galaxy having FWHM(H$_\beta$) $\lesssim1000$\rm~km~s$^{-1}$ \citep{zh06}. This AGN exhibits a relativistic jet aligned close to the line of sight \citep{yu08}. In this paper, we analyse archival \xmm{} (UV to hard X-ray), \suzaku{} (0.7$-$10\keV{}) data and study the spectral properties of the source, including the effect of partial covering absorption on the continuum as well as the disc/corona and jet emission. We apply different model-independent methods, including `flux-flux plot', `hardness-flux plot' and `RMS spectrum' to study the X-ray variability in the source.

We describe the \xmm{} and \suzaku{} observations and data reduction in Section~2. In Section~3, we present spectral fitting. In Section~4, we present timing analysis including the modelling of fractional RMS spectra. Finally, we discuss our results in Section~5. Throughout the paper, the cosmological parameters $H_0=73$\rm ~km~s$^{-1}$~Mpc$^{-1}$, $\Omega_{m}=0.27$, $\Omega_{\Lambda}=0.73$ are adopted. We have used the following convention to define power-law photon index: $f_{E} \propto E^{-\Gamma}$, where $f_{E}$ is the photon number flux density.

\section{Observation and Data Reduction}
\subsection{XMM-Newton Observations}
RXJ1633.3+4719 was observed four times with the \xmm{} telescope \citep{ja01}. The European Photon Imaging Camera (EPIC-pn) \citep{st01} was operated in the large window mode using the thin filter during all the observations. The first two observations were performed in 2011 (Obs. ID 0673270101 and 0673270201) with good exposure times of 20.13\ks{} and 22.77\ks{}, respectively. The next two observations were performed in 2012 (Obs. ID 0673270301 and 0673270401) with net exposure times of 15.14\ks{} and 10.96\ks{}, respectively. The data sets were processed with the Scientific Analysis System ({\tt SAS} v.14.0.0) and the most recent (as of 2014-12-25) calibration files. We processed only pn data using {\tt EPPROC} to produce calibrated photon event files. We have used unflagged single and double pixel events with {\tt PATTERN} $0-4$ to filter the processed pn events.

To exclude soft proton background flaring, we used filters on time by creating a GTI (Good Time Interval) file above 10\keV{} for the full field with RATE$<2.0$\rm~Counts~s$^{-1}$ using the task {\tt TABGTIGEN}. Then we extracted source spectra and light curves from filtered EPIC-pn event files with a source extraction radius of 25~arcsec. Similarly we extracted background spectra and light curves from nearby source free circular regions with radius of 50~arcsec. The source and background light curves for different energy bands and bin times were extracted using {\tt XSELECT} V.2.4 c from the cleaned event files. We produced background subtracted light curves using the {\tt FTOOL} task {\tt LCMATH}. We generated the RMF (Redistribution Matrix File) and ARF (Ancillary region File) for each EPIC-pn spectral data with the {\tt SAS} tasks {\tt RMFGEN} and {\tt ARFGEN}, respectively. Finally, we grouped the spectra using the {\tt GRPPHA} tool so that we had a minimum of 100 counts per energy channel.

The Optical Monitor (OM) was operated in the imaging mode using optical/UV (UVW2, UVM2, UVW1, U, B, V) filters in all four observations. We processed the OM data with the {\tt SAS} task {\tt OMICHAIN}. We identified the source in the sky aligned images in each filter. Then we examined the combined source list with the {\tt FTOOL} task {\tt FV} to find the RA, declination, count rate, magnitude and flux of the source. The OM response matrices for all filters and template filter files were obtained from the \xmm{} website\footnote[1]{\url{ftp://xmm.esac.esa.int/pub/ccf/constituents/extras/responses}}. Then we changed the `RATE' and `STAT-ERR' entries in the spectrum to the filter count rate and error, respectively.

\begin{table}
\caption{\xmm{} and \suzaku{} Observations of RXJ1633.3+4719}
\scalebox{0.75}{%
\begin{tabular}{cccccc}
\hline 
Observatory & Obs. ID & Date & Net Exposure & Net Count Rate \\
            &  &  & (ks)$^{a}$ & (counts s$^{-1}$)$^{b}$ \\                                      
\hline 
\xmm{} & 0673270101 & 2011-07-09 & 20.13 & (6.17$\pm0.06)\times10^{-1}$ \\ [0.2cm]
           & 0673270201 & 2011-09-12 & 22.77 & (6.24$\pm0.05)\times10^{-1}$ \\ [0.2cm] 
           & 0673270301 & 2012-01-14 & 15.14 & (7.17$\pm0.07)\times10^{-1}$ \\ [0.2cm]
           & 0673270401 & 2012-03-14 & 10.96 & (5.40$\pm0.07)\times10^{-1}$ \\ [0.2cm]
\hline 
\suzaku{} & 706027010 & 2011-07-01 & 79.98 & (7.50$\pm0.10)\times10^{-2}$ \\ [0.2cm]
      & 706027020 & 2011-07-18 & 75.02 & (7.12$\pm0.10)\times10^{-2}$ \\ [0.2cm]
      & 706027030 & 2012-01-13 & 88.25 & (9.30$\pm0.10)\times10^{-2}$ \\ [0.2cm]
      & 706027040 & 2012-02-05 & 90.97 & (6.30$\pm0.09)\times10^{-2}$ \\ [0.2cm]
\hline 
\end{tabular}}
NOTES: $^a$Net exposure times are quoted for \xmm{}/EPIC-pn and \suzaku{}/XIS 0+3. $^b$Count rates for EPIC-pn and XIS 0+3 are estimated in the 0.3$-$10\keV{} and 0.7$-$10\keV{} bands respectively. 
\label{table1}           
\end{table}

\begin{figure*}
\includegraphics[width=0.45\textwidth,angle=-0]{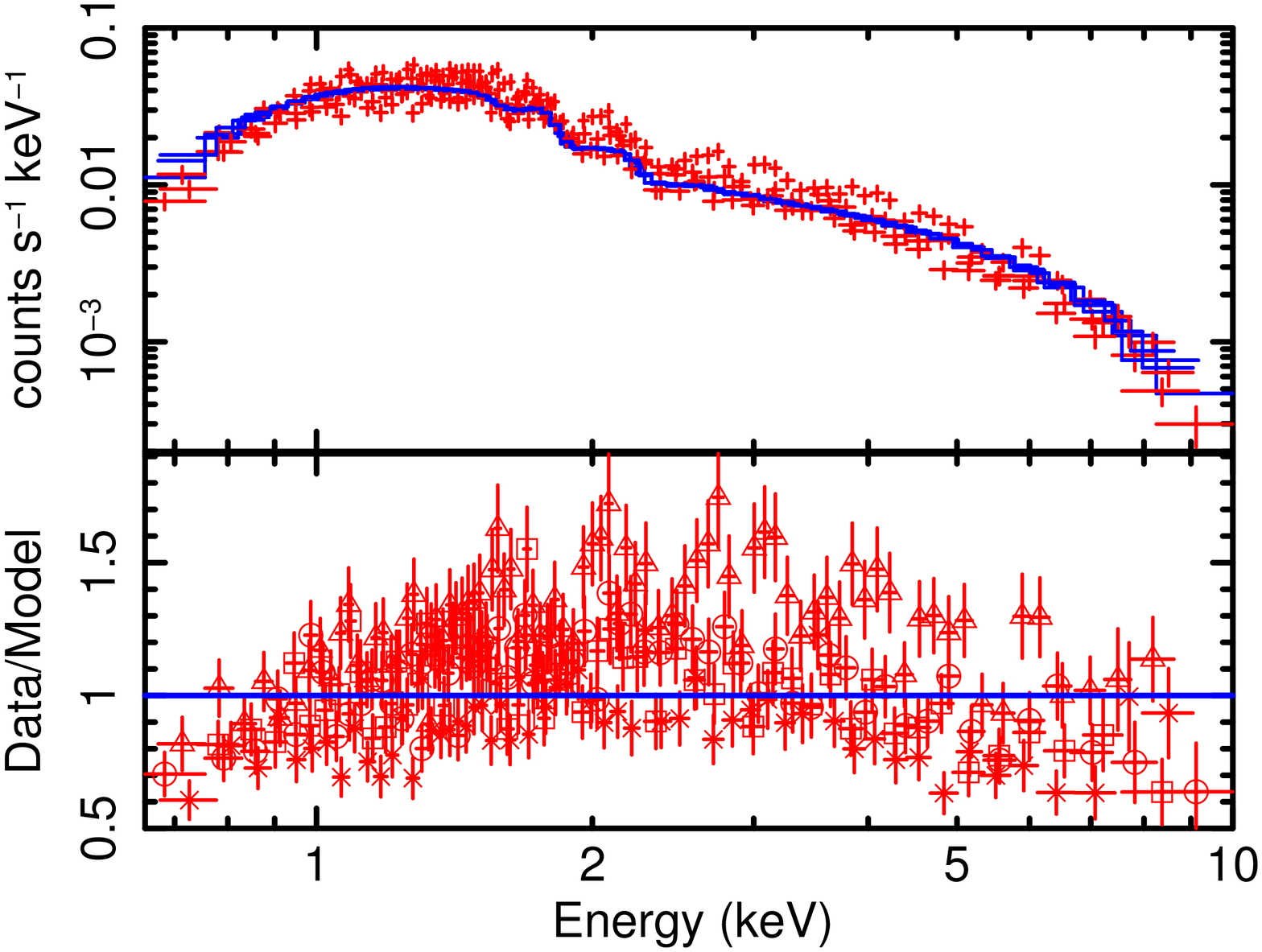}
\includegraphics[width=0.45\textwidth,angle=-0]{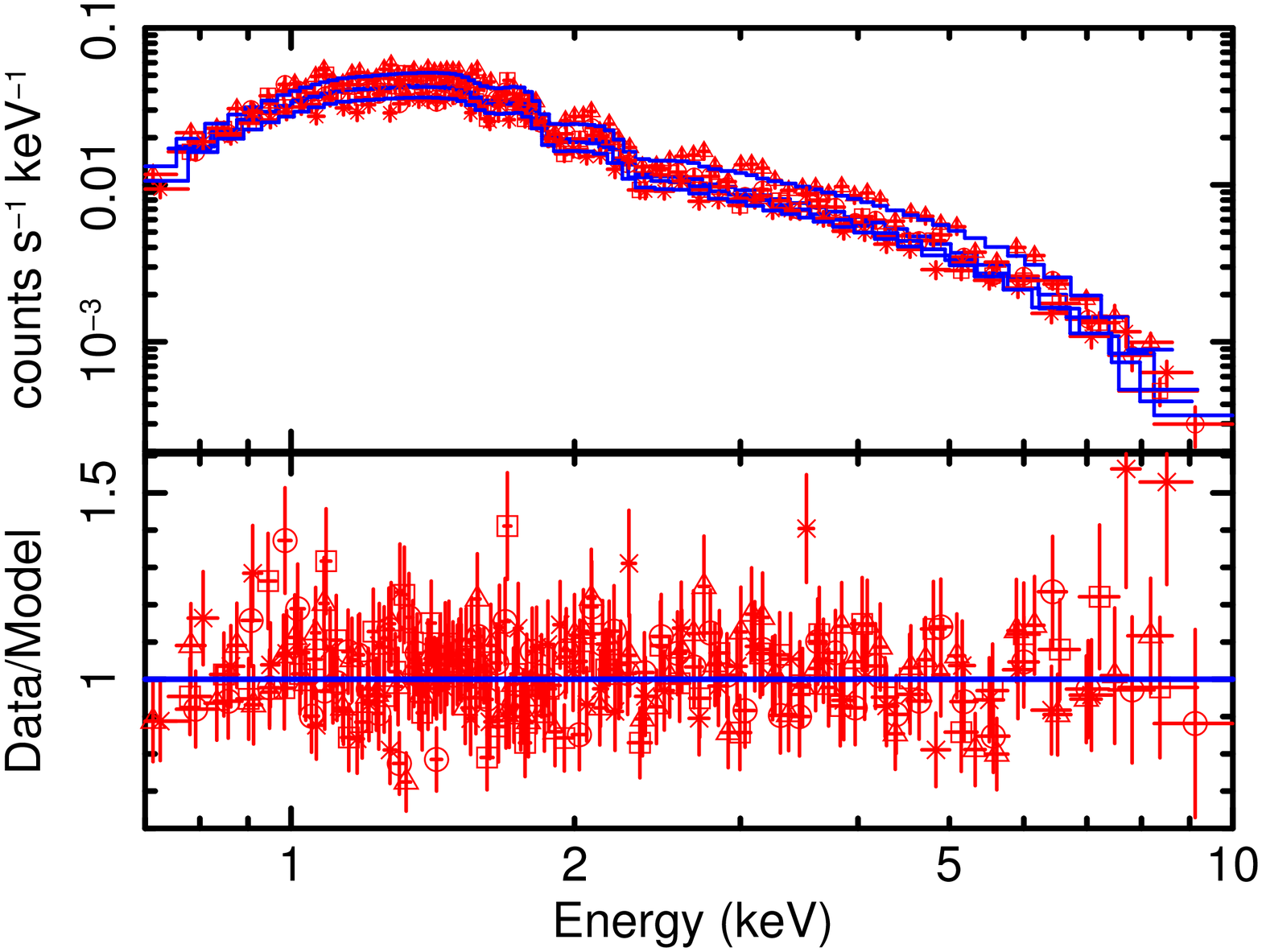}
\caption{{\it Left:} Four \suzaku{}/XIS~0+3 spectral data, the {\tt zpowerlw} model ($\Gamma = 1.8$) modified by the Galactic absorption and the ratio of data to model. {\it Right:} Four XIS~0+3 spectral datasets, the joint best-fitting model and the data-to-model ratio. The best-fitting model consists of a simple power-law modified by a partially covering, weakly ionized absorber along the line of sight. The circles, squares, triangles and crosses represent spectral data for Obs. ID 706027010, 706027020, 706027030 and 706027040, respectively.}
\label{fig1}
\end{figure*}

\begin{figure*}
\includegraphics[width=0.45\textwidth,angle=-0]{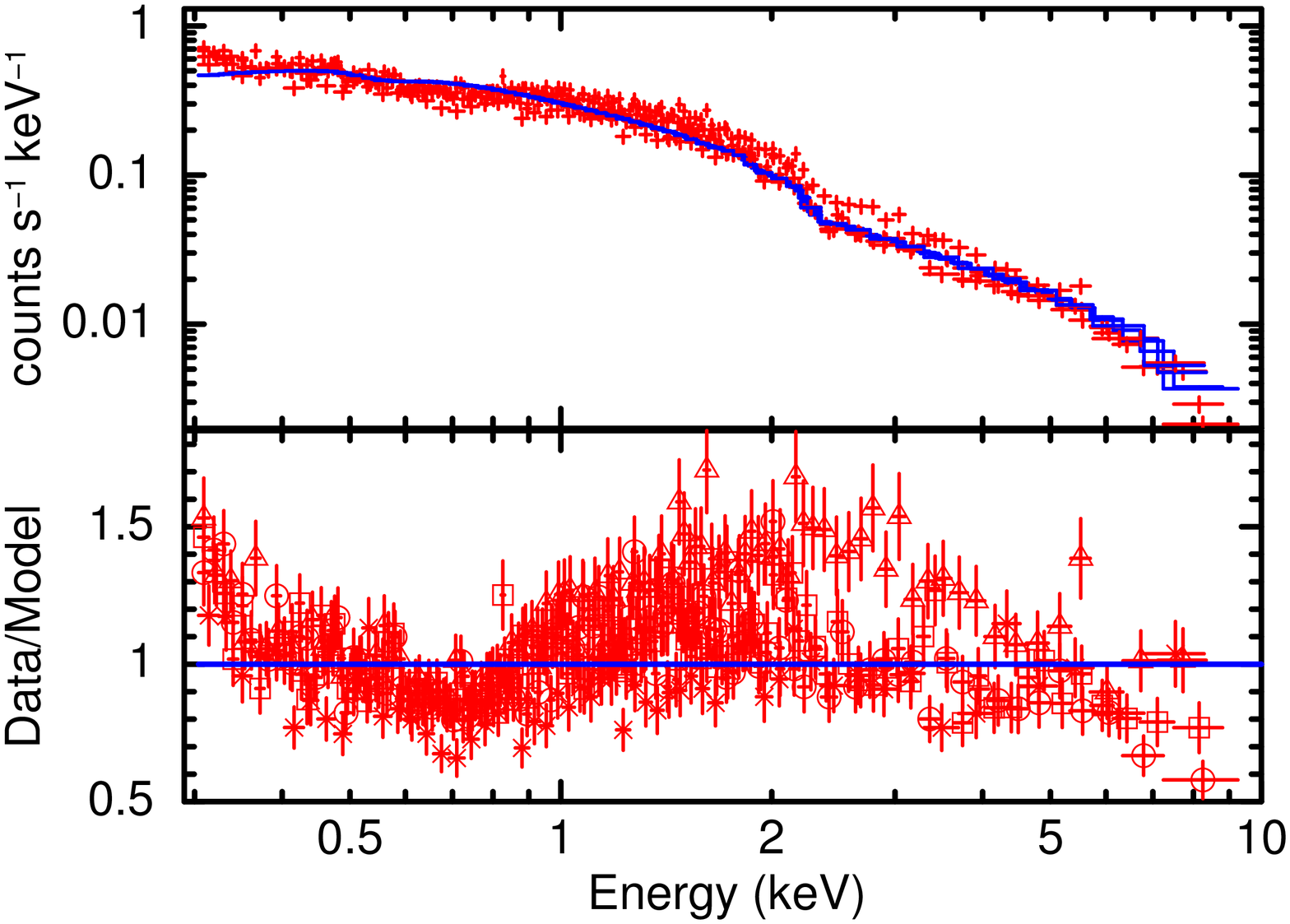}
\includegraphics[width=0.45\textwidth,angle=-0]{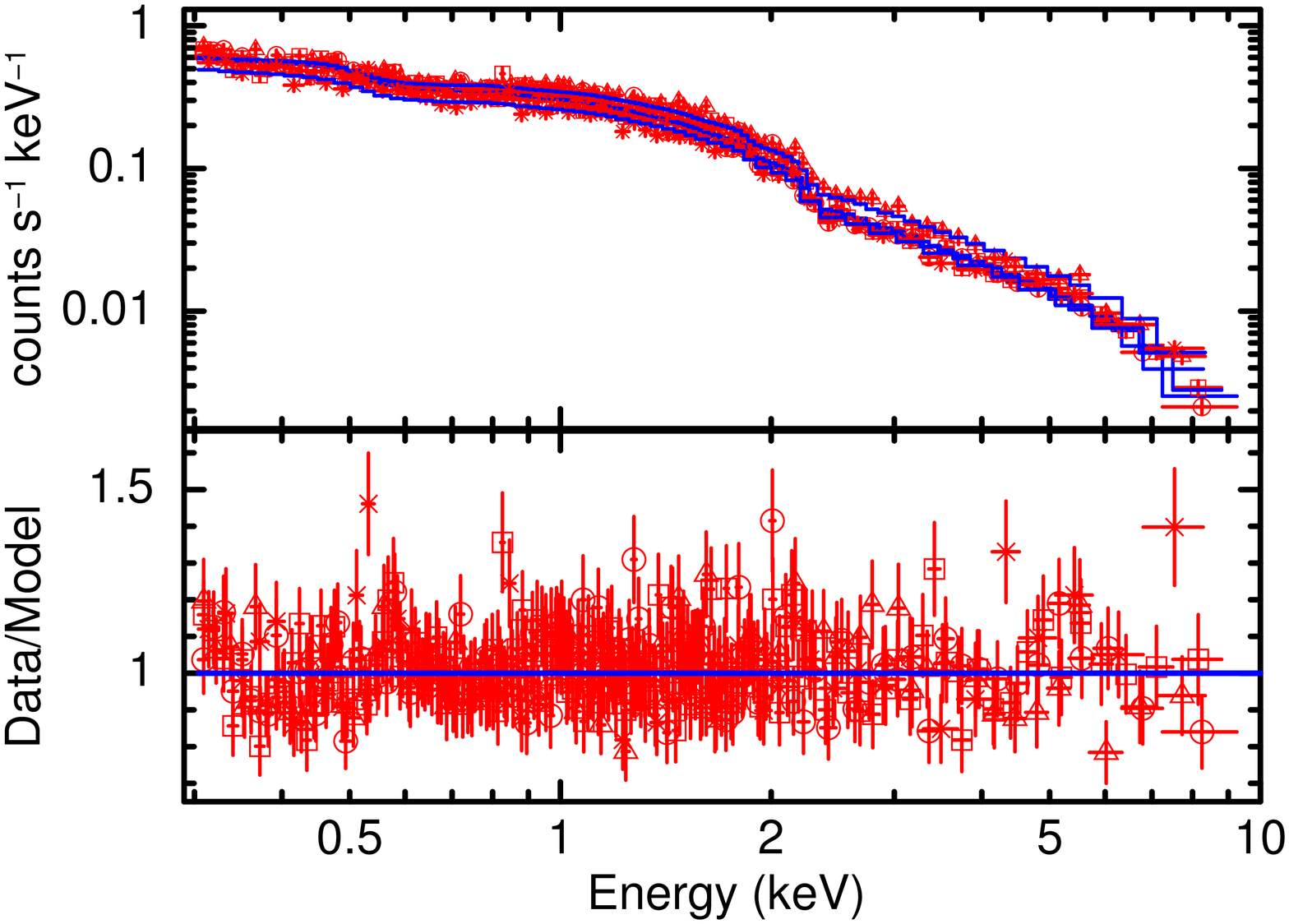}
\caption{{\it Left:} Four \xmm{}/EPIC-pn spectral data, the {\tt zpowerlw} model ($\Gamma = 1.68$) modified by the Galactic absorption and the ratio of data to model. {\it Right:} Four EPIC-pn spectral datasets, the joint best-fitting model and the data-to-model ratio. The best-fitting model consists of thermal Comptonization and blackbody spectrum of accretion disc modified by a partially/fully covering, weakly ionized absorber along the line of sight. The circles, squares, triangles and crosses indicate spectra for Obs. ID 0673270101, 0673270201, 0673270301 and 0673270401, respectively.}
\label{fig2}
\end{figure*}

\subsection{Suzaku Observations}
\suzaku{} has observed RXJ1633.3+4719 four times. There are four components of the X-ray Imaging Spectrometer (XIS): XIS0, XIS2, XIS3 are front-illuminated (FI) CCDs and XIS1 is back-illuminated CCD. However XIS2 is not in operation due to charge leaking in 2006 November. The other three CCDs (XIS0, XIS1, XIS3) were operated in 3$\times$3 and 5$\times$5 edit modes with normal clocking mode. The data sets were reduced using the software {\tt HEASOFT} v.6.16 and the recent CALDB (as of 2014-10-01) in conformity with the \suzaku{} {\tt ABC} guide (v.5.0). We reprocessed the XIS data with the task {\tt AEPIPELINE} to produce the cleaned event files, from which spectral products were extracted using {\tt XSELECT} V.2.4 c. Then we generated the XIS RMF (Redistribution Matrix File) and ARF (Ancillary Region File) using the tasks {\tt XISRMFGEN} and {\tt XISARFGEN}, respectively. We extracted the source spectra from a circular region of radius 120~arcsec centred on the source position and background spectra from a circular region of radius 150~arcsec free from both the source and the calibration sources inscribed in the corners of each CCD. The spectra and response files for all FI CCDs XIS~0+3 were combined using the tool {\tt ADDASCASPEC} in order to increase the S/N. Finally, we grouped the spectra using the {\tt GRPPHA} tool with a minimum of 100 counts per bin. 

\section{Spectral Analysis}
We analysed the spectra of RXJ1633.3+4719 obtained from the \xmm{} and \suzaku{} observations using {\tt XSPEC} v.12.8.2\citep{ar96}. We used the $\chi^{2}$ statistic and quote the errors at the $90\%$ confidence limit for a single parameter corresponding to $\Delta\chi^{2}$ = 2.704. 

\subsection{\suzaku{} XIS 0+3 spectra}
We analysed \suzaku{} 0.7$-$10\keV{} XIS 0+3 spectral data. Initially, we modelled the spectra with a continuum ({\tt zpowerlw}) modified by the Galactic absorption over the energy range 0.7$-$10\keV{}. We fixed the Galactic column density at $N_{\rm H}=1.71\times10^{20}$\rm~cm$^{-2}$ \citep{ka05}. This model provided an unacceptable fit with $\chi^{2}$=143.4 for 61 degrees of freedom (d.o.f) for Obs. ID 706027010. The deviations of the observed data from the {\tt zpowerlw} model are shown in Figure~\ref{fig1} ({\it Left}). We noticed some curvature in the energy range 1$-$5\keV{}, which may be caused by partial covering absorption (PCA). We multiplied a neutral PCA component {\tt zpcfabs} with the {\tt zpowerlw} model, which improved the fit from $\chi^{2}$/d.o.f = 143.4/61 to $\chi^{2}$/d.o.f = 64/59 ($\Delta\chi^{2}$=$-$79.4 for two additional parameters). To test whether the absorption is ionized, we replaced {\tt zpcfabs} by an ionized PCA component {\tt zxipcf} \citep{ree08}. This improved the fit from $\chi^{2}$/d.o.f = 143.4/61 to $\chi^{2}$/d.o.f = 61/58 ($\Delta\chi^{2}$=$-$82.4 for three additional parameters). Thus the ionized absorber model provided marginal improvement over the neutral absorption model. As we do not have the data below 0.7\keV{}, we were not able to investigate possible presence of a soft X-ray excess component. Our final model in the 0.7$-$10\keV{} band {\tt TBabs}$\times${\tt zxipcf}$\times${\tt zpowerlw} applied to the four observations resulted in $\chi^{2}$/d.o.f = 61.0/58, 62.8/51, 78.4/77, 55.5/54 for Obs. ID 706027010, 706027020, 706027030, 706027040, respectively. The best-fitting model parameters are listed in Table~\ref{table2}. We also performed a joint fit to the four XIS~0+3 spectral datasets which resulted in $\chi^{2}$/d.o.f=267.5/248. The XIS~0+3 spectral datasets, the best-fitting model and the deviations of the observed data from the best-fitting model are shown in Fig.~\ref{fig1} ({\it Right}).

\subsection{\xmm{} EPIC-pn/OM spectra}
We performed the EPIC-pn spectral analysis by fitting with a continuum ({\tt zpowerlw}) modified by the Galactic absorption over the energy range 0.3$-$10\keV{}. This resulted in a statistically poor fit ($\chi^{2}$/d.o.f=283.5/109 for Obs. ID 0673270101). The deviations of the observed data from the {\tt zpowerlw} model are shown in Figure~\ref{fig2} ({\it Left}). We noticed a surplus of emission below 0.5\keV{}, slight spectral curvature in the 1$-$5\keV{} band and an absorption feature $\sim0.7$\keV{} for Obs. Id 0673270401 (Fig.~\ref{fig2}, {\it Left}). Therefore we used the {\tt zxipcf} model \citep{ree08} to account for the spectral curvature and the deficit of emission near $\sim0.7$\keV{}. This improved the fit to $\chi^2$/d.o.f=127.0/106 ($\Delta\chi^{2}$=$-$156.5 for two additional parameters). Addition of a blackbody component ({\tt zbbody}) further improved the fit to $\chi^2$/d.o.f=110/104 ($\Delta\chi^{2}$=$-$17 for two additional parameters) with the temperature $kT_{\rm bb}$ = 26$^{+1.3}_{-0.1}$\ev{} which is in agreement with that found using \rosat{} PSPC data by \citet{yu10}. This blackbody temperature is a factor of $\sim4$ lower than the temperature of the soft X-ray excess ($kT\sim100-200$\ev{}) found for radio-quiet NLS1s and quasars with a large range of black hole masses \citep{cz03,gi04,cr06,de07}. While the temperature of the strong soft X-ray excess emission from NLS1 and quasars is too high to arise from a standard disc, the blackbody temperature of $\sim26$\ev{} for RXJ1633.3+4719 is entirely consistent with the expectations from an accretion disc in an AGN (see section~\ref{sec5.1}). Therefore, we use the accretion disc model below to describe the low temperature excess emission from RXJ1633.3+4719.

\begin{figure*}
\includegraphics[width=0.45\textwidth,angle=-0]{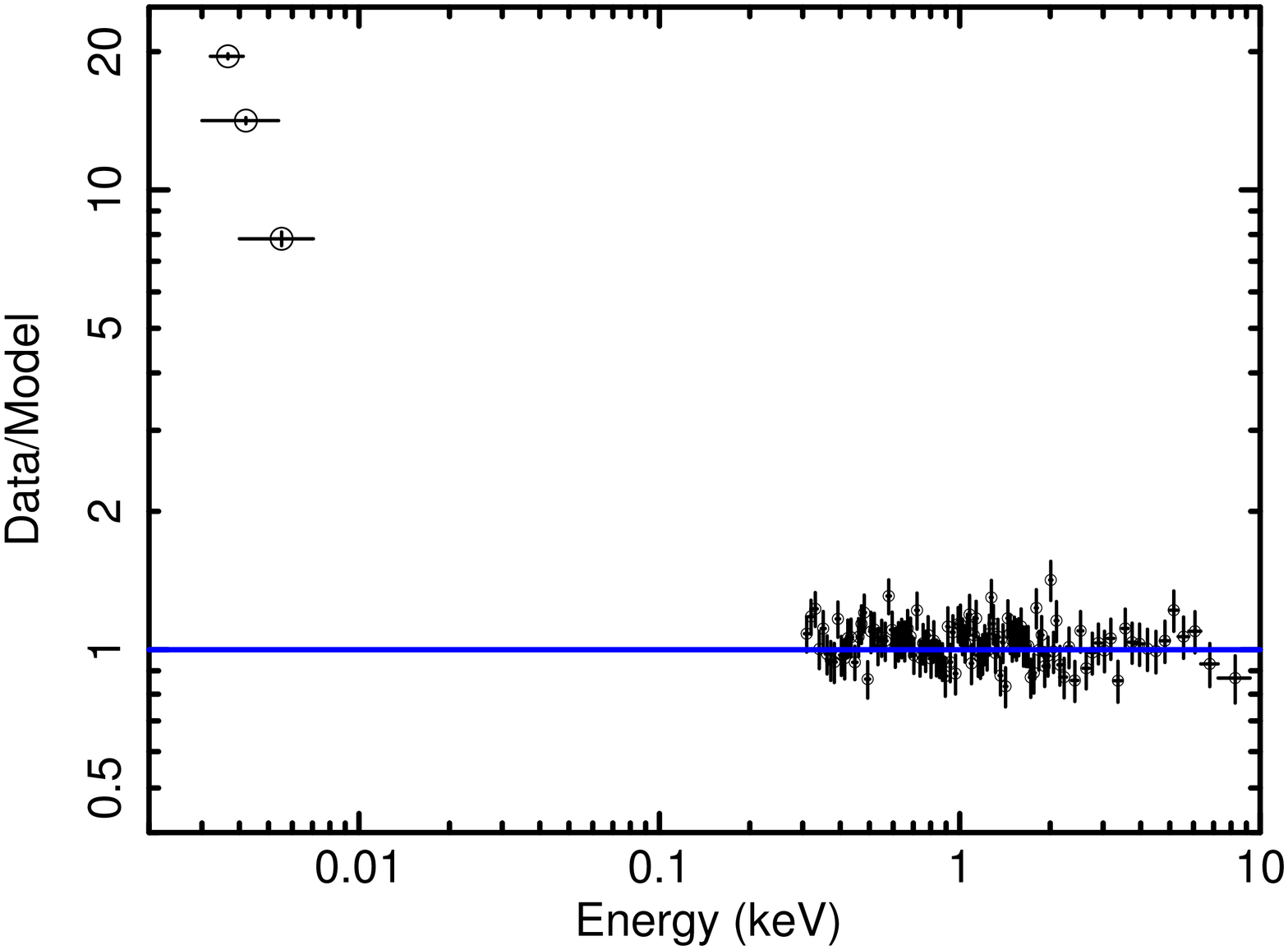}
\includegraphics[width=0.45\textwidth,angle=-0]{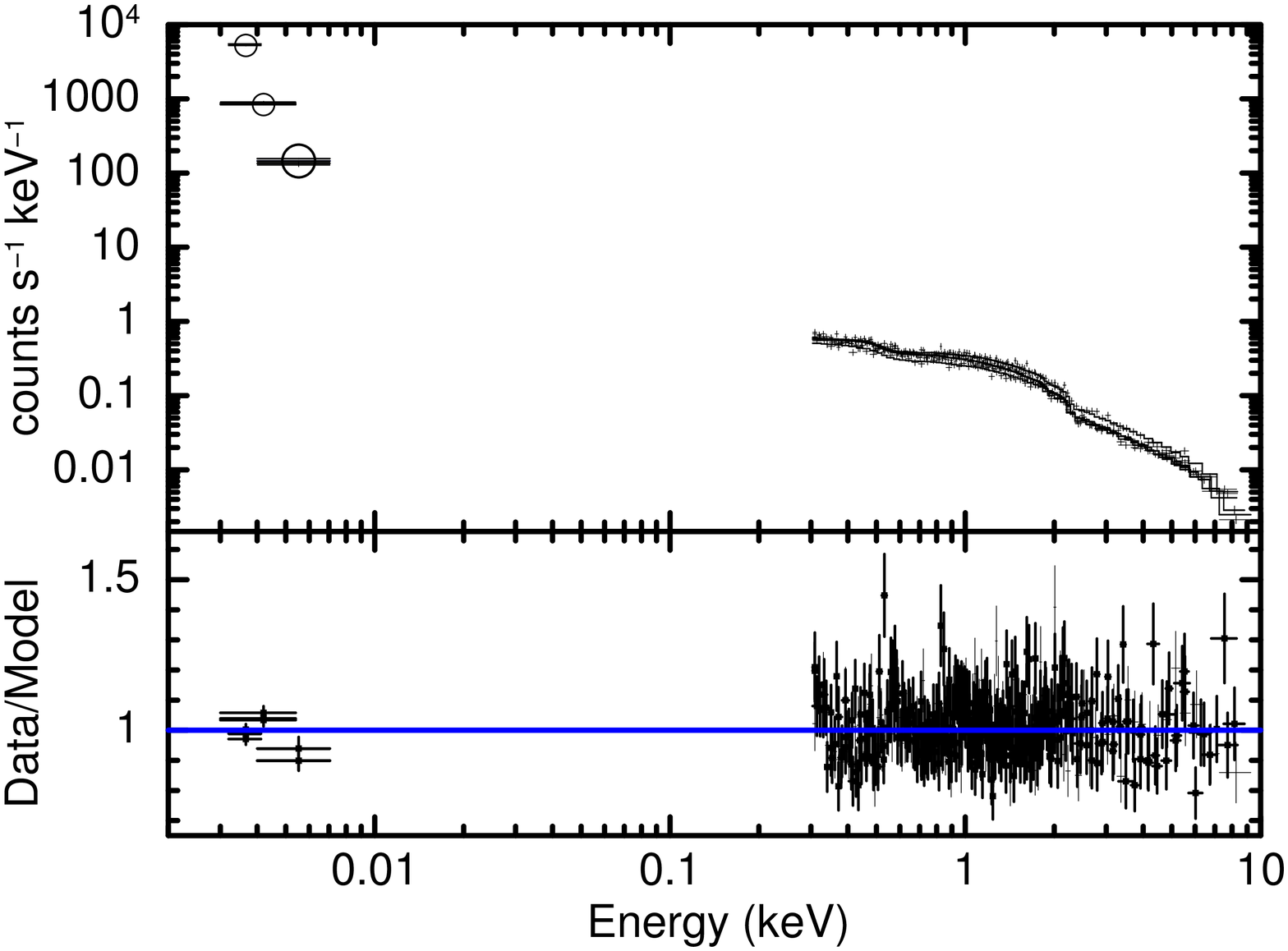}
\caption{{\it Left panel:} The ratio of OM (U, UVW1, UVM2), EPIC-pn data to the X-ray best-fitting model extrapolated down to UV band for Obs. ID 0673270101. {\it Right panel:} Four EPIC-pn and OM (U, UVW1, UVM2 filters) spectral datasets, the joint UV/X-ray best-fitting model and the data-to-model ratio. The best-fitting model consists of thermal Comptonization and blackbody spectrum of accretion disc modified by a partially/fully covering, weakly ionized absorber along the line of sight and an unabsorbed power-law.} 
\label{fig3}
\end{figure*}
 
The power-law emission is believed to be produced by the thermal Comptonization of soft disc photons in a hot corona with electron temperature $kT_{e}\sim$100 \keV{}. Therefore, we replaced {\tt zpowerlw} model by the thermal Comptonization model {\tt nthcomp} \citep{zd96,zy99}. To describe the excess emission below 0.5\keV{}, we have used the model {\tt diskpn} which provides soft seed photons for inverse Compton scattering in the hot electron corona \citep{gi99}. We fixed the inner disc radius at $r_{\rm in} = 6r_{\rm g}$ ($r_{\rm g}$ = GM/c$^2$). We tied the temperature of the seed photons in {\tt nthcomp} with the maximum temperature of the disc in the {\tt diskpn} component. The electron temperature of the thermal plasma was fixed at 100\keV{}. The {\tt diskpn} normalization is defined as: $n_{diskpn}=\frac{M^{2}cosi}{D^{2}\beta^{4}}$. We adopted SMBH mass $\sim3\times10^{6} M_{\odot}$ (\citealt{yu08, yu10}), disc inclination angle $\leq$10$^{\circ}$\citep{yu10} and color-to-effective temperature ratio $\beta$=2 \citep{ro92,sh95}. Therefore we fixed the {\tt diskpn} normalization at $2.06$. We also explored the possibility with $\beta=1$ (see e.g. \citealt{va09}). The best-fitting disc blackbody temperature is found to be 41.6$^{+2.5}_{-2.9}$\ev{} and 31.8$^{+1.4}_{-1.6}$\ev{} for $\beta=2$ and 1, respectively. 

\begin{figure}
\includegraphics[width=\columnwidth]{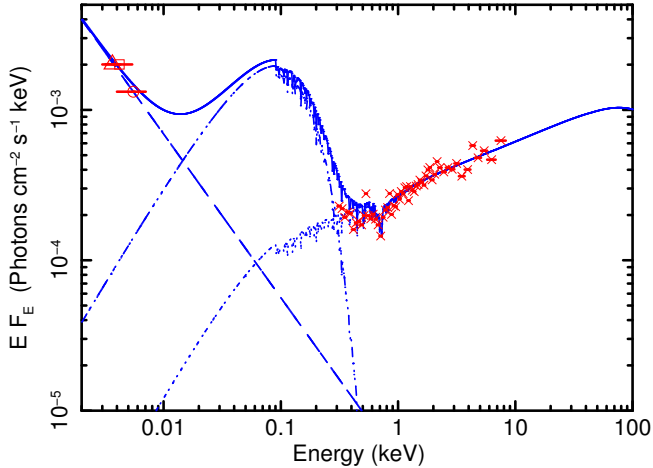}
\caption{Theoretical models used in this work (excluding the Galactic absorption and reddening components and including the intrinsic absorption component). The best-fitting model (solid line) has the following main components: the absorbed thermally Comptonized continuum (dotted), absorbed disc component (dotted-dashed) and unabsorbed power-law (dashed). The triangle, square, circle and crosses represent \xmm{} U, UVW1, UVM2 filters and EPIC-pn data respectively.}
\label{fig4}
\end{figure}

\begin{table*}
 \centering
 \caption{The best-fitting spectral model parameters for \suzaku{} observations (0.7$-$10\keV{}) of RXJ1633.3+4719. The best-fitting model in the 0.7$-$10\keV{} band consists of a power-law component ({\tt zpowerlw}) and an ionized absorption model ({\tt zxipcf}). {\tt TBabs} is used for the Galactic absorption. The notation `(f)' indicates that the parameter value was frozen. $^a$~Normalization in units of photons cm$^{-2}$ s$^{-1}$ keV$^{-1}$ at 1\keV{}. $^b$~Flux in the energy range 0.7$-$10\keV{} in units of erg~cm$^{-2}$~s$^{-1}$.}
\begin{center}
\scalebox{1.0}{%
\begin{tabular}{cccccc}
\hline 
Model & {\tt TBabs}$\times${\tt zxipcf}$\times${\tt zpowerlw} &  & \suzaku{} \\
\hline
 & Date & 2011-07-01 & 2011-07-18 & 2012-01-13 & 2012-02-05 \\ 
\hline
 & Obs. ID & 706027010 & 706027020 & 706027030 & 706027040 \\ 
\hline
Component & Parameter & &  &  & \\                                       
\hline
{\tt TBabs} & $N_{\rm H}$(10$^{20}$ cm$^{-2}$) & 1.71(f) & 1.71(f) & 1.71(f) & 1.71(f) \\
{\tt zxipcf} & $N_{\rm H}$(10$^{21}$cm$^{-2}$) & 6.3$^{+2.5}_{-4.2}$ & 5.1$^{+9.7}_{-4.5}$ & 13.8$^{+10.6}_{-8.3}$ & 20.6$^{+15.6}_{-8.1}$  \\ [0.2cm]
 & $\log\xi$(erg cm s$^{-1}$) & $-0.8^{+2.5}_{-p}$ & $-0.6^{+2.3}_{-p}$ & $-0.54^{+1.0}_{-0.7}$ & 1.1$^{+0.8}_{-0.6}$ \\ [0.2cm]
 & C$_f$(\%) & 75.8$^{+6.2}_{-12.3}$ & 67.6$^{+p}_{-20}$ & 70.7$^{+8.2}_{-11.0}$ & 72.7$^{+23.7}_{-12.6}$ \\ [0.2 cm]
{\tt zpowerlw} & $\Gamma$ & 2.3$^{+0.14}_{-0.15}$ & 2.2$^{+0.17}_{-0.15}$ & 2.33$^{+0.18}_{-0.19}$ & 2.38$^{+0.19}_{-0.18}$ \\ [0.2cm]
      & $n_{\rm PL}$(10$^{-4}$)$^{a}$ & 9.6$^{+2.5}_{-3.2}$ & 7.4$^{+2.3}_{-1.5}$ & 13.5$^{+4.8}_{-3.8}$ & 9.2$^{+3.7}_{-2.4}$ \\ [0.2cm]
FLUX & $f_{\rm PL}$(10$^{-12}$)$^{b}$ & 2.5$^{+0.3}_{-0.3}$ & 2.1$^{+0.3}_{-0.2}$ & 3.4$^{+0.5}_{-0.6}$ & 2.2$^{+0.3}_{-0.3}$ \\ [0.2cm]
     & $\chi^2$/$\nu$ & 61/58 & 62.8/51 & 78.4/77 & 55.5/54 \\
\hline
\end{tabular}}
\end{center}
\label{table2}
\end{table*}

\begin{table*}
 \centering
 \caption{The best-fitting spectral model parameters for \xmm{} observations (3\ev{}$-$10\keV{}) of RXJ1633.3+4719. The best-fitting model in the 0.3$-$10\keV{} band consists of a thermally Comptonized continuum ({\tt nthcomp}), blackbody spectrum of an accretion disc ({\tt diskpn}) and an ionized absorption model ({\tt zxipcf}). An unabsorbed power-law component ({\tt zpowerlw}) is required to fit the UV spectra. {\tt uvred} and {\tt TBabs} are used for Galactic extinction and absorption respectively. The notation `(f)' indicates that the parameter value was frozen. 
$^a$~Normalization in units of photons cm$^{-2}$ s$^{-1}$ keV$^{-1}$ at 1\keV{}. $^b$~{\tt diskpn} normalization in units of $n_{\rm diskpn}$=$\frac{M^{2}cosi}{D^{2}\beta^{4}}$, where $M$ is the central SMBH mass (in units of $M_{\odot}$), $D$ is the distance to the source (in units of \kpc{}), $i$ is inclination of the disc and $\beta$ is the color/effective temperature ratio.}
\begin{center}
\scalebox{0.9}{%
\begin{tabular}{cccccc}
\hline 
Model & {\tt uvred}$\times${\tt TBabs}$\times$[{\tt zpowerlw}$+${\tt zxipcf}({\tt diskpn}$+${\tt nthcomp})] &  &  & \xmm{} \\
\hline
 &Date& 2011-07-09 & 2011-09-12 & 2012-01-14 & 2012-03-14 \\                                       
\hline
 & Obs. ID & 0673270101 & 0673270201 & 0673270301 & 0673270401 \\ 
\hline
Component & Parameter &    \\
\hline 
{\tt uvred} & $E(B-V)$ & 0.018(f) & 0.018(f) & 0.018(f) & 0.018(f) \\
{\tt TBabs} & $N_{\rm H}$(10$^{20}$ cm$^{-2}$) & 1.71(f) & 1.71(f) & 1.71(f) & 1.71(f) \\
{\tt zxipcf} & $N_{\rm H}$(10$^{21}$cm$^{-2}$) & 4.4$^{+1.4}_{-1.4}$ & 3.9$^{+1.5}_{-1.6}$ & 6.7$^{+1.2}_{-3.5}$ & 1.4$^{+0.7}_{-0.7}$ \\ [0.2cm]
       & $\log\xi$(erg cm s$^{-1}$) & $-0.4^{+0.2}_{-0.6}$ & $-0.4^{+0.2}_{+0.4}$ & $-0.5^{+0.3}_{-1.5}$ & 0.9$^{+0.5}_{-0.7}$ \\ [0.2cm]
 & C$_f$(\%) & 69.2$^{+5.8}_{-6.5}$ & 64.1$^{+16.5}_{-6.9}$ & 73.3$^{+4.9}_{-4.9}$ & 100(f) \\ [0.2 cm]
{\tt diskpn} & $T_{\rm max}$(\ev{}) & 41.1$^{+13.9}_{-6.3}$ & 39.1$^{+2.6}_{-3.9}$ & 41.7$^{+8.5}_{-6.3}$ & 36.5$^{+19.9}_{-5.7}$ \\ [0.2cm]
       & $r_{\rm in}$($r_{\rm g}$) & 6(f) & 6(f) & 6(f) & 6(f) \\ [0.2cm]
       & $n_{\rm diskpn}^{b}$ & 2.2$^{+7.2}_{-2.1}$ & 2.06(f) & 2.4$^{+6.1}_{-1.8}$ & 2.6$^{+9.4}_{-2.5}$ \\ [0.2cm]
{\tt nthcomp} & $\Gamma$ & 2.09$^{+0.10}_{-0.08}$ & 1.95$^{+0.1}_{-0.05}$ & 2.10$^{+0.10}_{-0.11}$ & 1.71$^{+0.07}_{-0.06}$  \\ [0.2cm]
       & $T_{\rm bb}$(eV) & 41.1$^{+13.9}_{-6.3}$ & 39.1$^{+2.6}_{-3.9}$ & 41.7$^{+8.5}_{-6.3}$ & 36.5$^{+19.9}_{-5.7}$ \\ [0.2cm]
       & $n_{\rm nth}$(10$^{-4}$)$^{a}$ & 5.0$^{+0.8}_{-0.5}$ & 4.4$^{+0.7}_{-0.3}$ & 7.2$^{+1.2}_{-1.1}$ & 3.1$^{+0.3}_{-0.2}$ \\ [0.2cm]
{\tt zpowerlw} & $\Gamma_{\rm UV}$ & 3.05$^{+0.40}_{-0.20}$ & 2.81$^{+0.13}_{-0.09}$ & 3.25$^{+0.36}_{-0.25}$ & 3.10$^{+0.38}_{-0.22}$ \\ [0.2cm]
      & $n_{\rm PL}$(10$^{-6}$)$^{a}$ & 8.1$^{+16.6}_{-7.0}$ & 31.3$^{+21.5}_{-15.9}$ & 2.7$^{+4.7}_{-2.5}$ & 6.1$^{+14.9}_{-5.7}$ \\ [0.2cm]
     & $\chi^2$/$\nu$ & 125.4/105 & 150.3/121 & 91.2/93 & 61.6/52 \\ [0.2cm]
\hline
\end{tabular}}
\end{center}
\label{table3}
\end{table*}

The best-fitting values of the {\tt nthcomp} photon index, absorption column density and covering fraction are $\Gamma=2.09^{+0.10}_{-0.08}$, $N_{\rm H}=4.4^{+1.4}_{-1.4}\times10^{21}\rm~cm^{-2}$ and C$_f$(\%)=67.6$^{+5.4}_{-6.6}$, respectively. The best-fitting model {\tt TBabs}$\times${\tt zxipcf}({\tt diskpn}+{\tt nthcomp}) in the 0.3$-$10\keV{} energy band applied to the four \xmm{}/EPIC-pn observations resulted in $\chi^{2}$/d.o.f=113.53/105, 126.79/120, 77.98/93, 53.14/52 for Obs. ID 0673270101, 0673270201, 0673270301, 0673270401, respectively. We also performed a joint fit to the four EPIC-pn spectral datasets which resulted in $\chi^{2}$/d.o.f=407/381. The EPIC-pn spectral datasets, the best-fitting model and the deviations of the observed data from the best-fitting model are shown in Fig.~\ref{fig2} ({\it Right}). The addition of a blackbody disc component ({\tt diskpn}) to the thermal Comptonization model ({\tt nthcomp}) improved the fit from 450/382 to 407/381 ($\Delta\chi^{2}$=$-$43 for 1 additional free parameter) in the joint fit to the four EPIC-pn data and the estimated F-statistic value is 39.71 with null hypothesis probability of $8.16\times10^{-9}$. Thus the presence of the disc emission as a soft X-ray tail is $6.1\sigma$ significant as suggested by {\tt Ftest}.

We then included the OM/UV filters U ($\lambda_{\rm eff}\sim3440~\textrm{\AA}$), UVW1 ($\lambda_{\rm eff}\sim2910~\textrm{\AA}$), UVM2 ($\lambda_{\rm eff}\sim~2310\textrm{\AA}$) and used the reddening model {\tt uvred} to correct for the Galactic extinction. We fixed the color excess at $E(B-V)=0.018$ \citep{sc11}. First,  we attempted to model the UV data alone using {\tt diskpn} which provided large $\chi^{2}$/d.o.f=1060.6/2. Then we multiplied a reddening component {\tt zredden} to the {\tt diskpn} model in order to account for any intrinsic reddening. This resulted in poor fit with $\chi^{2}_{\rm red}=4.54$ for $\beta=2$ and $\chi^{2}_{\rm red}=4.51$ for $\beta=1$. We found unphysically high disc temperature $kT_{\rm max}\sim0.23$\keV{} and 80.3\ev{} for $\beta=2$ and 1, respectively with intrinsic reddening $E(B-V)\sim0.28$. Therefore the UV data cannot be explained by the reddened accretion disc model alone. 

We extrapolated the 0.3$-$10\keV{} X-ray best-fitting model to the UV band which resulted in strong UV excess emission with large $\chi^{2}$/d.o.f=9921/114 for Obs. ID 0673270101. The deviation of the observed data from the extrapolated X-ray best-fitting model is shown in Figure \ref{fig3} ({\it Left panel}) for Obs. ID 0673270101. The host galaxy contribution obtained from the SDSS optical spectrum of RXJ1633.3+4719 is $\sim20\%$ above $3600~\textrm{\AA}$ and negligibly small ($\sim3-4\%$) below $3600~\textrm{\AA}$ as shown in Figure 1 of \citet{yu08}. From that Figure it is also evident that the nuclear emission dominates the host galaxy contribution towards the UV band. The maximum percentage for the degree of host contamination in each OM/UV filter is estimated to be $\sim3\%$. RXJ1633.3+4719 is not a local AGN ($z=0.116$) and its UV luminosity is very high ($L\sim10^{44}$\rm~erg~s$^{-1}$), so it is expected that the AGN would outshine its host galaxy in the UV band. Since the source is a flat-spectrum, radio-loud AGN, there would be significant contribution of the jet in the UV band and we use a power-law to account for any contribution from the jet. As the jet is away from the torus, any UV emission from the jet is not affected by the intrinsic absorption. The addition of a power-law (without intrinsic absorption, {\tt zpowerlw}) to the X-ray best-fitting model can fit the broadband UV to hard X-ray data with $\chi^{2}$/d.o.f = 125.4/105. The best-fitting value of the {\tt zpowerlw} photon index is steep, $\Gamma_{\rm UV}$ = $3.05^{+0.40}_{-0.20}$.  Thus our final model from UV to hard X-ray band 
{\tt uvred}$\times${\tt TBabs}$\times$[{\tt zpowerlw}+{\tt zxipcf}({\tt diskpn}+{\tt nthcomp})] resulted in $\chi^{2}$/d.o.f = 125.4/105, 150.3/121, 91.2/93, 61.6/52 for Obs. ID 0673270101, 0673270201, 0673270301, 0673270401, respectively. The best-fitting model parameters are listed in Table \ref{table3}. We executed a joint fit to the four EPIC-pn and OM spectral datasets which resulted in $\chi^{2}$/d.o.f=456/379. The OM/EPIC-pn spectra and the deviations of the observed data from the best-fitting model are shown in Figure~\ref {fig3} ({\it Right panel}). The best-fitting model is shown in Figure~\ref {fig4}.

\section{Timing Analysis}
\begin{figure}
\includegraphics[width=\columnwidth]{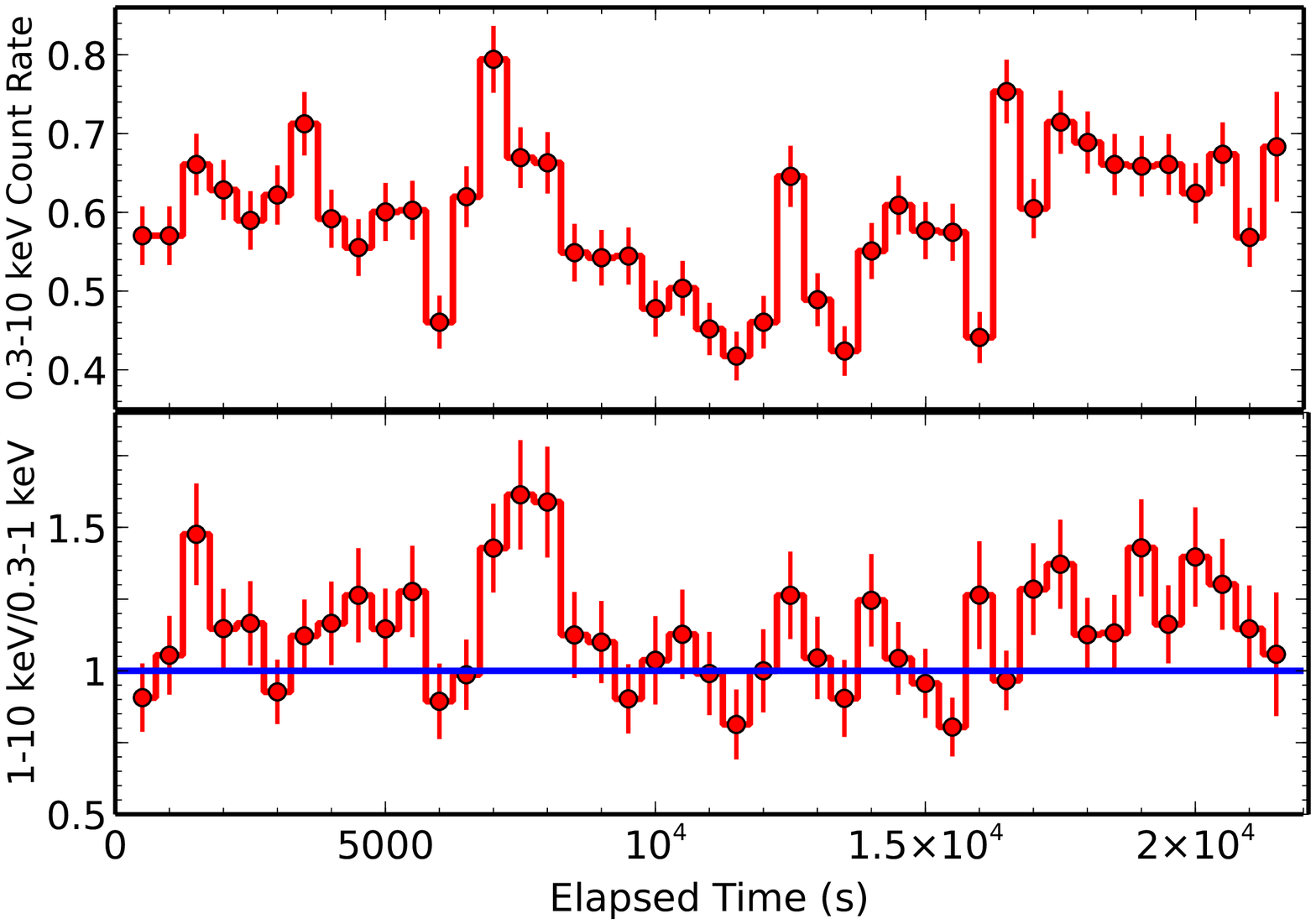}
\caption{{\it Upper panel:} \xmm{}/EPIC-pn (0.3$-$10\keV{}) background subtracted light curve with time bin size of 500\s{} for Obs. ID 0673270101. {\it Lower panel:} Hardness ratio (defined as the ratio of EPIC-pn count rate between 1$-$10\keV{} and 0.3$-$1\keV{} energy bands) as a function of time with bin size of 500\s{} for Obs. ID 0673270101. The solid blue line corresponds to the hardness ratio of 1. Most of the time hardness ratio is either greater than or equal to unity which means that the hard band (1$-$10\keV{}) flux dominates over the soft band (0.3$-$1\keV{}).}
\label{fig5}
\end{figure}

We study the energy dependent X-ray variability in the source by exploring different model-independent techniques. We extracted the source and background light curves in the full (0.3$-$10\keV{}), soft (0.3$-$1\keV) and hard (1$-$10\keV) bands using the same regions which were used for the spectral extraction. The upper panel of Figure~\ref{fig5} shows the background subtracted EPIC-pn light curve with $500\s$ bins. The source is variable with fractional variability amplitude $F_{\rm var}$=13.5$\pm1.0\%$ for Obs. ID 0673270101. The hard band variability amplitude ($F_{\rm var,hard}$=17.9$\pm1.4\%$) is large compared to that of the soft band ($F_{\rm var,soft}$=9.6$\pm1.7\%$), which indicates the possible presence of multiple spectral components varying differently. In the lower panel of Fig.~\ref {fig5}, we have shown the hardness ratio (defined as the ratio of count rate between 1$-$10\keV{} and 0.3$-$1\keV{} energy bands) as a function of time. For all four \xmm{} observations, the hardness ratio is either greater than or equal to unity which means that the hard band (1$-$10\keV{}) flux dominates over the soft band (0.3$-$1\keV{}) flux.

\begin{figure}
\includegraphics[width=\columnwidth]{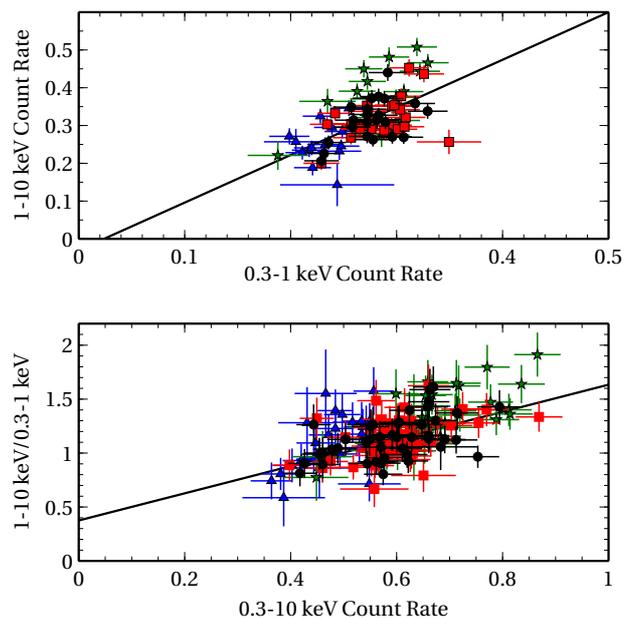}
\caption{{\it Upper panel:} The EPIC-pn count rate in the energy band 1$-$10\keV{} is plotted against the count rate in the energy band 0.3$-$1\keV{} with time bins of 1\ks{} for all four \xmm{}/EPIC-pn observations. The solid line shows the best linear fit to the data. The black circle, red square, green star and blue triangle represent data for Obs. ID 0673270101, 0673270201, 0673270301, 0673270401, respectively. The flux-flux plot exhibits significant scatter around the best-fitting line and a soft band (0.3$-$1\keV{}) offset. {\it Lower panel:} The hardness ratio (defined as the ratio of EPIC-pn count rate between 1$-$10\keV{} and 0.3$-$1\keV{} energy bands) is plotted as a function of 0.3$-$10\keV{} count rate for all four \xmm{}/EPIC-pn observations. The source becomes harder as the flux increases which is in contrast with radio-quiet AGN.}
\label{fig6}
\end{figure}

One way to investigate the flux/spectral variability is to study the flux-flux plot which represents the relationship between count rates in two different energy bands as introduced by \citet{ch01} and \citet{ta03}. Figure~\ref {fig6} ({\it Upper panel}) shows the variation of the EPIC-pn count rate in the 1$-$10\keV{} energy band as a function of the count rate in the 0.3$-$1\keV{} energy band. We fit the flux-flux plot with a linear relation of the form, $y=mx+c$, where $y$ and $x$ represent the 1$-$10\keV{} and 0.3$-$1\keV{} band count rates respectively. A linear fit to the data provided a $\chi^2$/d.o.f=492/78 with slope $m\sim1.26$ and soft offset $c\sim-0.03$ counts s$^{-1}$. The Spearman rank correlation coefficient between the soft and hard band flux is $\sim0.56$ with null hypothesis probability of $6.2\times10^{-8}$. We find significant scatter around the best-fitting line in the flux-flux plot which implies that the spectral variability in the soft and hard bands are different from each other.

In order to search for X-ray spectral variability, we explore the hardness ratio versus EPIC-pn count rate plot as shown in Figure~\ref {fig6} ({\it Lower panel}). The source becomes harder as the X-ray (0.3$-$10\keV{}) flux increases which can happen in two possible ways: the soft band (0.3$-$1\keV{}) flux may decrease or the hard band (1$-$10\keV{}) flux can increase. If the soft band flux decreases due to absorption, then the overall 0.3$-$10\keV{} count rate will also decrease, which is not the case here. So we can rule out absorption induced variability and conclude that only the primary continuum is responsible for the X-ray variability on observed time-scales.

\subsection{Fractional Variability or RMS Spectrum}
Another useful perspective is to look into the fractional RMS variability amplitude $F_{\rm var}$ as a function of energy or RMS spectrum. $F_{\rm var}$ is the square root of the normalized excess variance $\sigma_{\rm NXS}^2$ which by definition is the excess variance $\sigma_{\rm XS}^2$ divided by the square of the mean $\overline{x}$ \citep{va03}: 

\begin{equation}
 F_{\rm var}=\sqrt{\sigma_{\rm NXS}^2}=\sqrt{\frac{\sigma_{\rm XS}^{2}}{\overline{x}^2}}
\end{equation}
\begin{equation}
 \sigma_{\rm XS}^2=S^2-\overline{\sigma_{\rm err}^2}
\end{equation}
where $S^2$ is the sample variance and $\overline{\sigma_{\rm err}^2}$ is the mean square error which is defined by

\begin{equation}
\overline{\sigma_{\rm err}^{2}}=\frac{1}{N}\sum\limits_{i=1}^{N}\sigma_{\rm err,i}^{2}
\end{equation}
We derived the RMS variability spectra by generating background subtracted light curve with time bins of 500\s{} in different energy bands and then computed the fractional RMS variability in the light curve after subtracting the Poisson noise due to measurement errors. The RMS variability spectrum can be thought of as arising from variations in the spectral parameters. 

\begin{figure}
\includegraphics[width=1.1\columnwidth]{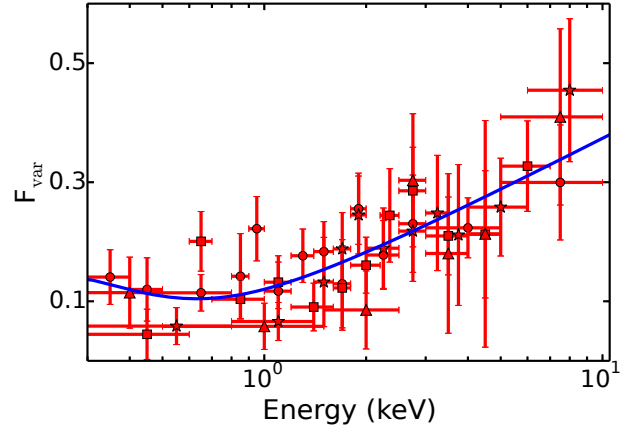}
\caption{The combined fractional RMS spectra of RXJ1633.3+4719 from four different \xmm{} observations of EPIC-pn data in the 0.3$-$10\keV{} energy band. The best-fitting model consists of a constant disc component and variable primary emission with the normalization and photon index anti-correlated. The solid blue line shows the best-fitting model to the data. The circle, square, star and triangle indicate data for Obs. ID 0673270101, 0673270201, 0673270301, 0673270401, respectively.}
\label{fig7}
\end{figure}

The mean X-ray spectrum of RXJ1633.3+4719 is well described by the combination of a disc component and a power-law, both modified by the Galactic absorption and intrinsic ionized absorption. The hardness-intensity diagram (Fig.~\ref {fig6}, {\it Lower panel}) suggested no significant absorption variability and the increasing $F_{\rm var}$ with energy (see Figure~\ref{fig7}) implies that dominant variability is due to the power-law component. If the observed X-ray variability is due to variation in the power-law normalization and slope while the disc component is steady, the fractional RMS can be written as:

\begin{equation}
 F_{\rm var}=\frac{\sqrt{<(\Delta f(A,\Gamma,E))^2>}}{f(A,\Gamma,E)}.
\end{equation}
where \begin{equation}
f(A,\Gamma,E)=f_{\rm PL}(A,\Gamma,E)+f_{\rm bbody}(E)= AE^{-\Gamma}+f_{\rm bbody}(E)
\end{equation}
Here $A$ and $\Gamma$ are the normalization constant and photon index of the power-law respectively. We then set

\begin{equation}
 F_{\rm{var}}=\frac{f_{\rm PL}(A,\Gamma,E)\sqrt{[\Delta\Gamma^{2}\log^{2}(E)+(\frac{\Delta A}{A})^{2}-2\alpha\frac{\Delta A}{A}\Delta\Gamma\log(E)]}}{f_{\rm PL}(A,\Gamma,E)+f_{\rm bbody}(E)}  
\label{eu1}
\end{equation}
where $\Delta \Gamma$ and $\Delta A$ are the variations in the slope and normalization of the power-law respectively, $\alpha$ is the correlation coefficient between the power-law normalization and index. 

We used the best-fitting parameters $\Gamma=2$, $A=6.5\times10^{-4}$ ph~cm$^{-2}$~s$^{-1}$~keV$^{-1}$ and $T_{\rm bb}=25$\ev{} obtained from the averaging of four energy spectral fitting and fit the combined fractional RMS spectra using the above model (equation \ref {eu1}). This model provided an acceptable fit with $\chi^{2}$/d.o.f = 35/40. The model is shown as a solid line in Fig.~\ref{fig7}. The variation in the normalization and photon index are $\frac{\Delta A}{A}\sim 12\%$ and $\frac{\Delta \Gamma}{\Gamma}\sim6.5\%$ respectively.

\begin{figure}
\includegraphics[width=0.46\textwidth,angle=-0]{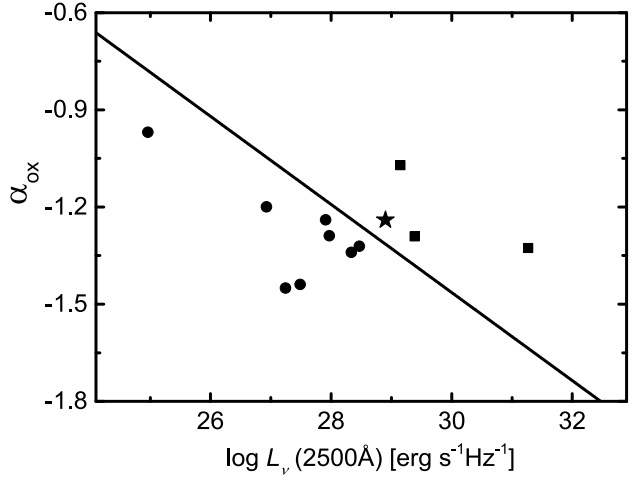}
\caption{Dependence of optical to X-ray spectral index, $\alpha_{\rm ox}$ on the monochromatic luminosity at $2500\textrm{\AA}$. The solid line represents the linear relation [$\alpha_{\rm ox}=-0.136\log L_{\nu}(2500\textrm{\AA})+2.616$] from \citet{str05}. The asterisk is for the source, RXJ1633.3+4719, squares are for radio-loud AGN from \citet{va09} and circles are for radio-quiet NLS1 galaxies from \citet{de08}.}
\label{alpha_ox}
\end{figure}

\begin{figure*}
\includegraphics[width=2\columnwidth]{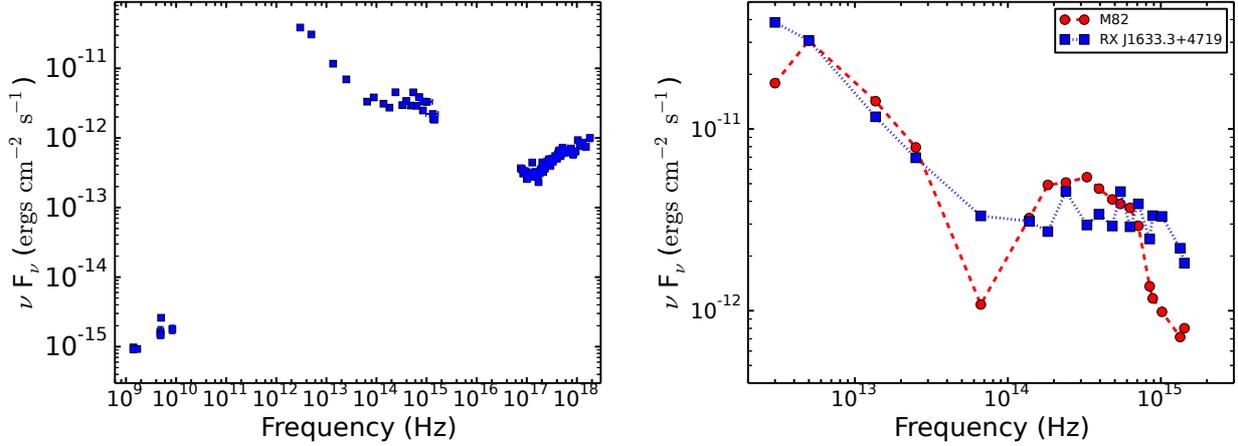}
\caption{{\it Left panel:} The radio-to-X-ray spectral energy distribution (SED) of RXJ1633.3+4719. The radio data were taken from the NASA/IPAC Extragalactic  Database (NED). {\it Right panel:} The infrared-to-UV SED of RXJ1633.3+4719 (blue square) and M82 (red circle) template normalized at 5$\times10^{12}$\hz{}. The NLS1 flux dominates over the M82 star-burst flux above 7$\times10^{14}$\hz{} i.e. $2.9$\ev{}. }
\label{fig9}
\end{figure*}

\section{Discussion}
In this paper, we study the broad-band (3\ev{}$-$10\keV{}) energy spectrum and the $0.3-10$\keV{} fractional RMS variability spectrum of the radio-loud NLS1 galaxy RXJ1633.3+4719 using four \xmm{} observations. Below we discuss the main results of this work. 

\subsection{Accretion disc emission}
\label{sec5.1} 
RXJ1633.3+4719 shows ultra-soft X-ray excess emission below $0.5\keV$. This excess is well described by a low temperature blackbody ($kT_{\rm bb}\sim25$\ev{}). For NLS1s and quasars the temperature of the soft X-ray component is much higher, $kT\sim100-200$\ev{} (see e.g. \citealt{cz03,gi04,cr06,de07}). In this regard, RXJ1633.3+4719 is different from other NLS1s. The best-fitting temperature of the blackbody component in RXJ1633.3+4719 is consistent with that expected from the standard accretion disc. We have derived a disc temperature of $\sim40$\ev{} for RXJ1633.3+4719 by using the disc model {\tt diskpn} for the soft excess. In case of a standard accretion disc, the local effective temperature \citep{sh73,no73} at radius $r$ is given by

\begin{equation}
T(r)\approx 54~\Big(\frac{r}{r_{\rm s}}\Big)^{-3/4}\Big(\frac{M_{\rm BH}}{10^{8}M_{\odot}}\Big)^{-1/4}\Big(\frac{\dot{M}}{\dot{M}_{\rm Edd}}\Big)^{1/4} \ev{}
\end{equation}
where $r$ is the distance from the central SMBH, $M_{\rm BH}$ is the mass of the central black hole in units of $M_{\odot}$ and $\frac{\dot{M}}{\dot{M}_{\rm Edd}}$ is the scaled mass accretion rate. 
When a black hole of mass $10^{8}M_{\odot}$ accretes matter at its Eddington limit, the standard disc temperature is expected to be $\sim24$\ev{} for a non-rotating black hole and $\sim50$\ev{} for a maximally rotating Kerr black hole. 
The Eddington luminosity can be written as 

\begin{equation}
L_{\rm Edd} \approx 1.38\times10^{38}\left(\frac{M_{\rm BH}}{M_{\odot}}\right) \rm~erg~s^{-1}
\end{equation}
For $M_{\rm BH}\sim 3\times10^{6} M_{\odot}$ \citep{yu08, yu10}, the Eddington luminosity is $L_{\rm Edd}=4.14\times10^{44}\rm~erg~s^{-1}$. We obtained the bolometric luminosity $L_{\rm bol}\approx 1.51\times10^{44}\rm~erg~s^{-1}$ by calculating the unabsorbed flux in the energy band $0.001-100$\keV{} using the convolution model {\tt cflux} in {\tt XSPEC}. Thus the Eddington ratio or scaled mass accretion rate for this radio-loud AGN is $\dot{m}_{\rm Edd}=\frac{\dot{M}}{\dot{M}_{\rm Edd}}=\frac{L_{\rm bol}}{L_{\rm Edd}}\approx1.5/4.1\approx0.37$. The accretion disc temperature becomes maximum at $r=\frac{49}{36}\times r_{\rm in}$, where $r_{\rm in}$ is the inner disc radius. For a non-spinning black hole, $r_{\rm in}=3r_{\rm s}$. Therefore the disc temperature reaches its peak value $T_{\rm max}$ at $r= \frac{49}{12}\times r_{\rm s}$ and the expression for the peak disc temperature is 

\begin{equation}
T_{\rm max}\approx 18.9~\Big(\frac{M_{\rm BH}}{10^{8}M_{\odot}}\Big)^{-1/4}\Big(\frac{\dot{M}}{\dot{M}_{\rm Edd}}\Big)^{1/4} \ev{}.
\label{eq2}
\end{equation}  
For $\dot{m}_{\rm Edd}=\frac{\dot{M}}{\dot{M}_{\rm Edd}}\approx0.37$ and $M_{\rm BH}\sim3\times10^{6} M_{\odot}$, Eqn. \ref {eq2} gives $T_{\rm max}\approx 35$\ev{}, which is close to the best-fitting value of the maximum disc temperature of $\sim40$\ev{}. 

\begin{figure}
\includegraphics[width=0.95\columnwidth]{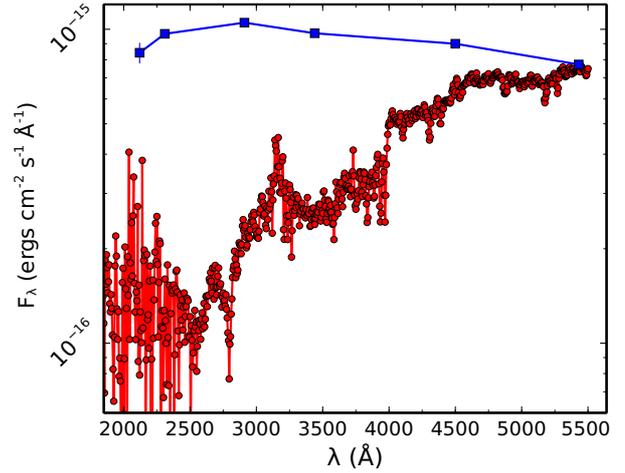}
\caption{The SED of RXJ1633.3+4719 (blue square) and normal spiral galaxy (Sb) template (red circle) normalized at $5445~\textrm{\AA}$.}
\label{fig10}
\end{figure}

\subsection{Origin of UV emission}
As shown above, the ultra-soft excess emission from RXJ1633.3+4719 is consistent with an accretion disc origin. However, the accretion disc component cannot explain the excess emission in UV and we need an additional steep power-law for the UV continuum. We have examined the strongest emission lines present in the OM bands--UVM2 ($1970-2675~\textrm{\AA}$), UVW1($2410-3565~\textrm{\AA}$) and U ($3030-3890~\textrm{\AA}$). From the composite quasar spectrum \citep{va01}, we find that the total equivalent widths of the strongest emission lines in the UVM2, UVW1 and U bands are $<50~\textrm{\AA}$, $<60~\textrm{\AA}$ and $<25~\textrm{\AA}$ respectively, which are much narrower than the widths of the broadband OM UV filters. Therefore, the excess UV emission from RXJ1633.3+4719 cannot be due to the contribution of the emission lines. 

To understand whether the UV emission (OM bandpass) is from the accretion disc or jet, we estimate the optical-to-X-ray spectral index $\alpha_{\rm ox}$ which is defined as

\begin{equation}
\alpha_{\rm ox}=-0.386\log\bigg[\frac{L_{\nu}(2500~\textrm{\AA)}}{L_{\nu}(2\keV{})}\bigg]
\end{equation}
where $L_{\nu}(2500~\textrm{\AA)}$ and $L_{\nu}(2\keV{})$ are the monochromatic luminosity at $2500~\textrm{\AA}$ and 2\keV{} respectively. We fit a power-law $f_{\nu}\propto\nu^{-\alpha}$ to calculate the monochromatic flux $f_{\nu}$ at any desired frequency $\nu$, 

\begin{equation}
f_{\nu}=\frac{h(1+\alpha)E^{\alpha}F}{E_{2}^{(1+\alpha)}-E_{1}^{(1+\alpha)}}
\end{equation}
where $h=4.138\times10^{-18}$ is the Planck constant in \keV{}-sec, $\alpha$ is the spectral index, $E_{1}$ and $E_{2}$ are respectively the lower and upper limits of energy over which flux density is integrated to get the broadband flux $F$. The optical-to-X-ray spectral index of the source is $-1.24$, shown as an asterisk in Figure~\ref{alpha_ox} and is consistent with $\alpha_{\rm ox}$ values for radio-loud AGN (3C~390.3, 3C~120 and 3C~273) taken from \citet{va09}.

To further investigate whether the excess UV flux is caused by possible presence of a nuclear star-burst in the host galaxy, we constructed the spectral energy distribution (SED) of RXJ1633.3+4719 using data from {\it 2MASS}, {\it WISE}, {\it IRAS} 60 and 100$~\rm {\mu m}$\footnote[2]{\url{http://irsa.ipac.caltech.edu/applications/Gator/}} and compared it with the SED of M82, which is one of the best studied star-burst galaxies. The rest-frame $1.4$\ghz{} luminosity of M82 ($z=0.00068$) is $L_{\rm 1.4\ghz{}}\sim10^{22}\rm~W\hz{}^{-1}$, which is two order of magnitude smaller than that of RXJ1633.3+4719. Figure~\ref{fig9} shows the SED of RXJ1633.3+4719 and M82. The SEDs were normalized at 5$\times10^{12}$\hz{} ($60~\rm{\mu m}$) where the star-burst contribution peaks. The flux from the AGN dominates over the M82 flux above 7$\times10^{14}$\hz{} i.e. $2.9$\ev{}. Thus we can conclude that the UV flux of RXJ1633.3+4719 is not affected by a putative nuclear star-burst. 

The best diagnostic to distinguish between star-forming galaxies and radio-loud AGN is the FIR-radio correlation parameter $q$ defined by 

\begin{equation}
q=\log\bigg[\frac{S_{\rm FIR}/(3.75\times10^{12})}{S_{\rm 1.4\ghz{}}}\bigg]
\end{equation}
where $S_{\rm 1.4\ghz}$ is the 1.4\ghz{} flux density in units of $\rm~W~m^{-2}~\hz{}^{-1}$ and $S_{\rm FIR}$ is the far-infrared flux density in units of $\rm~W~m^{-2}$ defined by

\begin{equation}
S_{\rm FIR}=1.26\times10^{-14}\bigg(2.58\times S_{\rm 60\mu m}+S_{\rm 100\mu m}\bigg).
\end{equation}
The star-burst galaxies have $q>1.8$. The estimated $q$ parameter for RXJ1633.3+4719 is 1.2 which is much lower than the cut-off of $q=1.8$.

In order to examine whether the strong UV excess emission is due to the host galaxy contamination in the OM bandpass, we have taken one normal spiral galaxy (Sb) template from \citet{ki96}\footnote[3]{\url{ftp://ftp.stsci.edu/cdbs/grid/kc96/}} and compared its SED with our source RXJ1633.3+4719. The host template was normalized to the source flux at $5445~\textrm{\AA}$. We found that the template SED rises steeply with wavelength while the SED of RXJ1633.3+4719 is flat ($F_{\lambda}\propto{\lambda}^{0.3}$). The SED of RXJ1633.3+4719 and Sb template are shown in Figure~\ref{fig10}. We have also shown that the maximum percentage for the degree of host galaxy contamination in UV is $\sim3\%$. Therefore the excess UV emission cannot be attributed to the host contamination. The only possibility left then is the high energy tail of synchrotron emission from a jet, which can be described by a power-law as observed. The slope of the UV spectrum ($\Gamma_{\rm UV}\approx3$) is consistent with the slope for flat spectrum radio quasars (FSRQ) and originates from the high energy end of the synchrotron component. Further investigations using $\gamma$-ray data from Fermi-Large Area Telescope would be useful (see e.g. \citealt{pa14, pa15}) to fully test the jet scenario in RXJ1633.3+4719.

\begin{figure*}
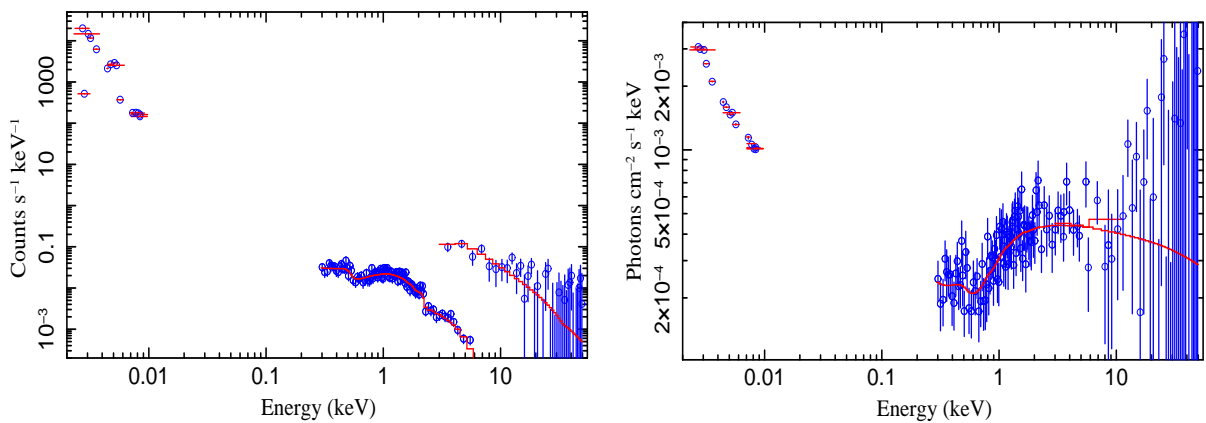

\includegraphics[width=5.6cm,height=8cm,angle=-90]{fig11a.ps}
\includegraphics[width=5.6cm,height=8cm,angle=-90]{fig11b.ps}
\caption{Simulated 50\ks{} \astr{} UVIT ($1300-5300~\textrm{\AA}$), SXT ($0.3-8$\keV{}) and LAXPC ($3-50$\keV{}) spectra (folded and unfolded in the left and right panels respectively) of RXJ1633.3+4719 using the best-fitting model {\tt uvred}$\times${\tt TBabs}$\times$[{\tt zpowerlw}+{\tt zxipcf}({\tt diskpn}+{\tt nthcomp})] inferred from the \xmm{}/OM and EPIC-pn data.}
\label{fig11}
\end{figure*}

\subsection{Broadband Continuum} 
The broadband UV to hard X-ray spectral fitting for RXJ1633.3+4719 reveals the presence of jet emission in the UV, disc emission as ultra-soft X-ray excess below 0.5\keV{}, partial covering ionized absorption below 1\keV{} and variable primary emission in the hard X-ray band above 2\keV{}. We did not find any reflection signature including the Fe line or strong soft X-ray excess in the energy spectra. The spectral modelling using \suzaku{} XIS data in the $0.7-10$\keV{} band agrees well with the modelling of \xmm{} EPIC-pn data. Although the disc extends down to the innermost regions as implied by the ultra-soft excess emission, the absence of blurred reflection suggests lack of strong illumination of the disc by the hot corona. This could be possible if the corona acts as the base of the jet \citep{fe99,ma05} and the coronal emission is beamed away from the disc.

From the modelling of the broadband spectrum, we estimated the relative contribution of different components to the observed flux in UV and X-ray bands. The UV luminosity integrated from $1700~\textrm{\AA}$ to $4000~\textrm{\AA}$, covering three UV filters U, UVW1, UVM2, is $L_{\rm UV}=L_{\rm UV,disc+jet}\approx7.9\times10^{43}$\rm~erg~s$^{-1}$. The jet luminosity $L_{\rm UV,jet}\approx7.2\times10^{43}$\rm~erg~s$^{-1}$ is $\sim90\%$ of $L_{\rm UV}$, considering negligible host galaxy starlight contribution below $4000~\textrm{\AA}$ \citep{yu08}. The X-ray (0.3$-$10\keV{}) luminosity is $L_{\rm X}\approx6.7\times10^{43}$\rm~erg~s$^{-1}$ where the jet component contributes only $1.5\%$ with $L_{\rm X,jet}\approx 10^{42}$\rm~erg~s$^{-1}$. Thus the jet contribution to the X-ray band is negligible compared to the disc/corona emission.

\subsection{Steady disc and variable corona}
The X-ray emission from RXJ1633.3+4719 is variable with fractional variability amplitude $F_{\rm var}$=13.5$\pm1.0\%$ and the variability is energy dependent (see Fig.~\ref{fig5} and Fig.~\ref{fig7}). The fractional variability amplitude is nearly constant at energies below 1\keV{} and increases with energy above 1\keV{} (see Fig.~\ref{fig7}). The soft band (0.3$-$1\keV{}) offset in the flux-flux plot and constant $F_{\rm var}$ below 1\keV{} indicate the presence of a soft constant component similar to that observed in IRAS~13224-3809 \citep{ka15}. 

The source hardness increases with increasing 0.3$-$10\keV{} X-ray flux which is in contrast with radio-quiet AGN \citep{vau01,pa07,em11}. The hardening of the source with increasing 0.3$-$10\keV{} X-ray flux rules out absorption-induced variability and allows us to conclude that the X-ray spectral variability in RXJ1633.3+4719 is predominantly due to the primary continuum which may vary both in normalization and spectral shape. The modelling of the X-ray variability spectrum as a constant disc component and variable primary emission explains the observed fractional RMS(E) quite well. We find that the normalization and photon index of the primary continuum are anti-correlated with each other by $50\pm{16}\%$ with 1$\sigma$ significance. Such X-ray spectral variability cannot be due to variations in the Comptonization seed photons alone, because that would lead to spectral steepening with flux. The dominant X-ray variability of RXJ1633.3+4719 must be intrinsic to the hot corona which may be related to the dynamic base of the jet. Simultaneous multiwavelength observations with \astr{} will be very useful to study disc/corona and jet connection in AGN. In Figure~\ref{fig11}, we have shown the simulated broadband energy spectra of RXJ1633.3+4719 using our best-fitting model and response matrices of UVIT, SXT and LAXPC instruments onboard \astr{} with 50\ks{} exposure time.

\section{Acknowledgments}
We thank the anonymous referee for constructive comments which helped to improve the manuscript significantly. LM is thankful to Chris Done, Mayukh Pahari and Raghunathan Srianand for their useful comments and discussions. This project is partially funded by UK-India UKIERI/UGC Thematic partnership grant UGC 2014-15/02 with University of Southampton, UK. This research has made use of archival data of \xmm{} and \suzaku{} observatories through the High Energy Astrophysics Science Archive Research Center Online Service, provided by the NASA Goddard Space Flight Center.

\bsp	
\label{lastpage}
\end{document}